\documentstyle[11pt,epsfig,axodraw]{article} 
\def\baselinestretch{1.3}
\newcommand{\ba}{\begin{array}}
\newcommand{\ea}{\end{array}}
\newcommand{\bd}{\begin{displaymath}}
\newcommand{\ed}{\end{displaymath}}
\newcommand{\be}{\begin{equation}}
\newcommand{\ee}{\end{equation}}
\newcommand{\bea}{\begin{eqnarray}}
\newcommand{\eea}{\end{eqnarray}}

\def\bra{\langle}
\def\ket{\rangle}

\def\e{\epsilon}

\def\l{\lambda}
\def\m{\mu}
\def\n{\nu}

\def\q2 {q^2}

\begin{document}
\begin{flushright}
{\large MRI-PHY/P990307 \\March, 1999  \\ hep-ph/9903418}
\end{flushright}

\begin{center}
{\Large\bf Radiative decay of the lightest neutralino in an R-parity
violating supersymmetric theory}\\[20mm]
Biswarup Mukhopadhyaya and Sourov Roy\\
{\em Mehta 
Research Institute,\\
Chhatnag Road, Jhusi, Allahabad - 211 019, India} 
\\[10mm]
\end{center}
\begin{abstract}
In an R-parity violating supersymmetric scenario, the lightest
neutralino $\tilde \chi^0_1$ is  no longer a stable particle. We 
calculate the branching ratio for the decay mode $\tilde \chi^0_1 
\longrightarrow \nu \gamma$ which occurs at the one-loop level. 
Taking into account bilinear as well as trilinear lepton number 
violating interactions as the sources of R-parity violation, we 
make a detailed scan of the parameter space, both with and without 
gaugino mass unification and including the constraints on the 
neutrino sector from the recent Superkamiokande results. 
This study enables one to suggest interesting experimental signals 
distinguishing between the two types of R-parity breaking, and also 
to ascertain whether such radiative decays can give rise to collider 
signals of the type $\gamma \gamma$ + $\not {\rm E}$ from 
pair-produced neutralinos. 
\end{abstract}

\vskip 1 true cm

\noindent
PACS NOS. : 12.60.Jv, 13.10.+q, 14.80.Ly

\newpage
\setcounter{footnote}{0}

\def\baselinestretch{1.8}
\section{Introduction}

Decay of a heavy particle with a photon in
the final state  can often lead to experimental signals for 
physics beyond the Standard Model.  Recent literature contains many 
studies, both experimental and theoretical,
devoted to signals of the type $\gamma \gamma$ + $\not {\rm E}$, $\gamma$ 
+ $\not {\rm E}$ etc. \cite{1,2} in the context of various new physics 
scenarios. 
A rather large fraction of such studies are concerned with supersymmetric
(SUSY) \cite{3} theories. 
For example, in models with
gauge mediated supersymmetry breaking (GMSB)\cite{4,5}, the decay of 
the lightest neutralino ($\tilde \chi^0_1$) (next-to-lightest 
supersymmetric particle  in these models) into a photon and a gravitino  
can  give rise to $\gamma \gamma$ + $\not {\rm E}$ (or $\gamma$ + 
$\not {\rm E}$)  which can have substantial event rates even after subtracting
the Standard Model backgrounds. The signal $\gamma \gamma$ + $\not {\rm E}$ 
is in fact the most probable discovery channel of GMSB. Side by side,  
radiative decay of the second lightest neutralino($\tilde \chi^0_2$),
in a scenario based on N = 1 supergravity (SUGRA), 
into a photon and the lightest neutralino ($\tilde \chi^0_1$, the lightest 
supersymmetric particle in this class of models) can sometimes give rise to
similar final states in colliders from a pair of $\tilde \chi^0_2$'s or 
from a ${\tilde \chi^0_2}{\tilde \chi^0_1}$ pair. It requires a careful 
investigation to distinguish the latter type of signals from the former to
establish the reality of GMSB vis-a-vis the SUGRA-type mode of
supersymmetry breaking \cite{6}.

In view of this, it is important to know whether there can be other 
new physics options which allow radiative decay of a heavier particle 
into a photon and another invisible particle, leading to signals of the same
kind as those discussed above, and, in some cases, faking the GMSB signals
which are being looked for so carefully. It will be obvious that the 
closest kinematical resemblance to the GMSB signal is borne when the 
invisible particle is
a near-massless neutral fermion, for which the immediate candidate is a 
neutrino. Now,  radiative decay of a heavier neutral fermion  
into a neutrino becomes a distinct possiblity in R-parity violating 
SUSY \cite{7} where the lightest neutralino can decay in this 
vein at the one-loop level. This motivates one to undertake an 
exhaustive study of the radiative decay 
$\tilde \chi^0_1 \longrightarrow \nu \gamma$ in an R-parity violating 
scenario, to see whether in some region of the parameter space this kind of a
decay can have an appreciable branching ratio in spite of the 
allowed tree-level decay channels of the lightest neutralino. In this paper,
we have attempted such a study, taking into account R-parity violation
through both bilinear and trilinear terms in the superpotential. 

It should be mentioned that $\tilde \chi^0_1 \longrightarrow \nu \gamma$ 
has been calculated in some earlier works \cite{8} where either the 
composition of the lightest neutralino has been confined to certain limits, 
or some specific regions of the SUSY parameter space have been adhered to. 
Our purpose, on the other hand, is to make a  detailed scan of the 
parameter space, and the results presented by us highlight those particular 
regions where the branching ratio can be of the largest magnitude. This, 
we feel, is necessary to establish the authenticity of, for example, 
GMSB signals during collider searches. We also go beyond the assumption 
of gaugino mass unification at the scale of grand unified theories (GUT) 
and present some results with the $SU(2)$ and $U(1)$ gaugino masses 
treated as independent parameters, to probe whether radiative neutralino 
decays with enhanced rates ensue upon freeing the SUSY standard model 
from a GUT embedding. And lastly, we take into account the fact that 
R-parity violating SUSY is being invoked in recent times to explain the 
generation of neutrino masses \cite{9} in consonance with the 
Superkamiokande data on atmospheric neutrinos \cite{10}. Thus a section 
of our results pertains to that particular region of the parameter space 
where the indicated  hierarchy of neutrino masses, together with large 
angle mixing between the second  and the third generations, is reproduced.

The paper is organised as follows. In section 2, we describe the 
basic framework of the R-parity violating model used in our calculations.
In section 3, some 
details of the loop calculation are presented. Section 4 contains 
our numerical results for various scenarios as well as for different
choices of the parameters.  We summarise and conclude in section 5 .
Detailed forms of the loop integrals are outlined in the appendix.

\section{Basic Framework}

The fact that all the particles in the standard model
carrying baryon (B) and lepton (L) number are fermions, with fixed gauge 
multiplet stuctures, implies that B and/or  L  cannot be violated by one
unit. Thus one cannot write down
renormalizable terms in the Lagrangian which violate B or L or both.
This is no longer true in the case of supersymmetric theories where the
particle spectrum is now doubled and baryon number and lepton number are
assigned to the supermultiplets. One can now have scalar particles
containing B or L, and it is possible to have  terms with 
$\Delta L$ or $\Delta B =1$. However, the scenario has to be made 
consistent with the non-observation of  proton decay which requires the 
simultaneous violation of B and L. This is ensured in an
over-restrictive manner in the minimal supersymmetric standard model (MSSM), 
by imposing a discrete
multiplicative symmetry called R-parity defined as $R = (-1)^{L + 3B + 2S}$,
which equals +1 for Standard Model particles and -1 for the
superpartners. An immediate consequence of R-parity conservation is that 
the lightest supersymmetric particle (LSP) is stable.

The conservation of R-parity, however, is not prompted by any strong 
theoretical reason, and theories where R is violated through nonconservation
of {\it either} B {\it or} L are perfectly consistent with stability of 
the proton. Such scenarios can be studied by genaralising  the MSSM 
superpotential to the following form :  

\begin{equation}
W = W_{MSSM} + W_{\not R}
\end{equation}
with
\begin{equation}
W_{MSSM} = {\mu} {\hat H}_1 {\hat H}_2 + h_{ij}^l {\hat L}_i {\hat
H}_1 {\hat E}_j^c
+ h_{ij}^d {\hat Q}_i {\hat H}_1 {\hat D}_j^c + h_{ij}^u {\hat Q}_i
{\hat H}_2 {\hat U}_j^c
\end{equation}
and
\begin{equation}
W_{\not R} = \lambda_{ijk} {\hat L}_i {\hat L}_j {\hat E}_k^c +
\lambda_{ijk}' {\hat L}_i {\hat Q}_j {\hat D}_k^c + 
\lambda_{ijk}''{\hat U}_i^c {\hat D}_j^c {\hat D}_k^c + \epsilon_i {\hat
L}_i {\hat H}_2
\end{equation}

In this paper we shall assume that the B-violating piece $\lambda''$ is 
not there. 
Also, we discuss the L-violating terms in two separate categories 
for the convenience of analysis, considering, in turn,  $W_{\not R}$ with
either the bilinear ($\epsilon_i L_i H_2$) \cite{11,12,13} or the trilinear   
($\lambda$- and $\lambda'$) \cite{14} terms existing in the superpotential 
at a time.

\underline {Case (1).}  $W_{\not R} = \epsilon_3 {\hat L}_3 {\hat H}_2$

We simplify the analysis here by assuming a bilinear R-parity 
violating term involving only the third leptonic generation. 
All the imporatnt features of such a scenario can be seen within
simplified framework.  

At first sight it appears 
that the $\epsilon$-term can be rotated away from the superpotential 
by properly redefining the $L_3$ and $H_1$ fields, so that

\begin{equation}
{{\hat H}'}_1 = \frac {\mu {\hat H}_1 + \epsilon_3 {\hat L}_3} 
{\sqrt{\mu^2 + \epsilon_3^2}}
\end{equation}

\begin{equation}
{{\hat L}'}_3 = \frac {-\epsilon_3 {\hat H}_1 + \mu {\hat L}_3} 
{\sqrt{\mu^2 + \epsilon_3^2}}
\end{equation}
Now as a result of this rotation the superpotential takes the form:

\begin{eqnarray}
W & = & {\mu'} {{\hat H}'}_1 {\hat H}_2 + h_{33}^{\tau} {{\hat L}'}_3
{{\hat H}'}_1 {\hat E}_3^c
+ \frac {h_{ij}^d \epsilon} {\mu'} {\hat Q}_i {{\hat H}'}_1 {\hat D}_j^c 
+ h_{ij}^u {\hat Q}_i {\hat H}_2 {\hat U}_j^c \nonumber
   \\[1.5ex]
& - & \frac {h_{ij}^d \epsilon} {\mu'} {\hat Q}_i {{\hat L}'}_3 {\hat D}_j^c 
- \frac {h_{ij}^l \epsilon} {\mu'} {\hat L}_i {{\hat L}'}_3 {\hat E}_j^c 
+ \frac {h_{ij}^l \mu} {\mu'} {\hat L}_i {{\hat H}'}_1 {\hat E}_j^c 
\end{eqnarray}
\noindent
which means that the $\lambda$- and $\lambda'$ type terms are generated
in general and in the last two terms $i = 1,2$.

But more importantly, it should be noted that the scalar potential has 
terms of the following forms in the original basis:

\begin{equation}
V_{scal} = M^2_{L_3} {\tilde L}^2_3 + m^2_1 H^2_1 + B_1 \mu H_1 H_2 +
B_2 \epsilon {\tilde L}_3 H_2 + \mu \epsilon_3 {\tilde L}_3 H_1 + ........
\end{equation}
where we have written down the F-term and soft breaking terms relevant
for our purpose here. It is evident that such terms will generate, in
general, a vacuum expectation value (vev) $v_3$ for the sneutrino. 
The basis rotation regenerates such terms, and therefore the sneutrino 
vev is in general non-vanishing as the special consequence of bilinear 
R-parity violation whenever soft SUSY breaking is there \cite{15,16}.

Thus there are two ways of parametrizing a bilinear R-parity
violating scenario: (i) with $\epsilon$-term present in the superpotential
together with a vev of the sneutrino, and (ii) with  $\epsilon$-term rotated
away from the superpotential but the vev of the sneutrino (in the rotated 
basis) embodying the L-violating effects that the former would imply in, say,
the neutralino and chargino mass matrices. Of course, the parameter
$\epsilon$ takes refuge in this case in the scalar potential together 
with the soft breaking parameter $B_2$. $B_2$ can be eliminated
by using the conditions for electroweak symmetry breaking, if
$\epsilon$ and $v_3$ are used as independent variables. Our calculations here
are done in a  basis where both $\epsilon$
and $v_3$ (vev of the sneutrino) are present in the superpotential.

Now, the presence of $\epsilon$ and $v_3$ will induce a kind of mixing in the
fermionic as well as in the scalar sector of the theory, which is typical of
bilinear R-parity violation. In the
fermionic sector neutralinos will mix with the tau-neutrino and the
charginos, with the tau lepton. Consequently, the
neutralino mass matrix takes the following form:

\begin{equation}
{{\cal M}_{\tilde \chi^0_1}} =  \left( \begin{array}{ccccc}
  0 & -\mu & \frac {gv} {\sqrt{2}} & 
  -\frac {g'v} {\sqrt{2}} & 0 \\
  -\mu & 0 & -\frac {gv'} {\sqrt{2}} 
       & \frac {g'v'} {\sqrt{2}} & 0 \\
 \frac {gv} {\sqrt{2}} & -\frac {gv'} {\sqrt{2}} & M & 0 & -\frac {gv_3} 
 {\sqrt{2}} \\
 -\frac {g'v} {\sqrt{2}} & \frac {g'v'} {\sqrt{2}} & 0 & M' & 
  \frac {g'v_3} {\sqrt {2}} \\
 0 & 0 & -\frac {gv_3} {\sqrt {2}} & \frac {g'v_3} {\sqrt {2}} & 
 0 
 \end{array}  \right)    
\end{equation}                                 

\noindent
where the successive rows and columns correspond to
(${\tilde H}_2, {\tilde H}_1, -i\tilde{W_3}, 
-i\tilde{B}, \nu_\tau$).
Here 

$$
v\ \ (v') = \sqrt{2}\ {\left(\frac {m^2_Z} {\bar{g}^2} 
 - \frac {v_3^2} {2} \right)}
^{\frac {1} {2}} {{\sin} \beta}\ \ ({{\cos} \beta})
$$

\noindent
$M$ and $M'$ are the ${\rm SU(2)}$ and ${\rm U(1)}$ 
gaugino mass parameters 
respectively, $\mu$, 
the Higgsino mass parameter,
and $\bar{g}=\sqrt{g^2+{g'}^2}.$  Evidently, this will result in the
generation of (Majorana) masses for the third generation neutrino at tree-level
through a see-saw type mechanism \cite{17}.  

In such a situation, the $(3{\times}3)$ chargino mass matrix is
\begin{equation}
M_{{\tilde \chi}^{\pm}} = \left(\matrix{M & -g {v_2} & 0\cr
  -g {v_1} & \mu & f {v_3}\cr
    -g {v_3} & \e & -f{v_1}\cr}\right)
    \end{equation}
    where $v_1 = \bra{H_1}\ket, ~~v_2 = \bra{H_2}\ket,
    ~~v_3 = \bra{{\tilde \nu}_{\tau}}\ket$, $\tan\beta = \frac {v_2}
    {v_1}$ and  
    ~$f = h^{l}_{33} = {\frac {m_{\tau}}{v_1}}$,
    ~$M$ being the $\rm SU(2)$ gaugino mass parameter. Here we have
    assigned
    $(-i{\bar {{\tilde W}^-}}, \bar {{{\tilde H}_1}^-}, {\bar
    {{\tau}_L}^-})$
    along the rows and
    $(-i{\bar {{\tilde W}^+}}, \bar {{{\tilde H}_2}^+}, {\bar
    {{\tau}_R}^+})$
    along the columns. 

Similarly, in the scalar sector neutral and charged scalar mass matrices
are enlarged as a result of the mixing between charged sleptons and
charged Higgs and sneutrino with the neutral Higgs. In the neutral
scalar sector the scalar mass-squared matrix in the basis $\{Re(H_1),
Re(H_2), Re(\tilde \nu_\tau)\}$ takes the form:

\begin{equation}
M_{s}^2 = \left(\matrix
{m_{1}^2+2{\l}c+4{\l}v_{1}^2 & -4{\l}{v_1}{v_2}+{B_1}{\m} &
4{\l}{v_1}{v_3}+{\m}{\e}\cr
-4{\l}{v_1}{v_2}+{B_1}{\m} & m_{2}^2-2{\l}c+4{\l}v_{2}^2 &
-4{\l}{v_3}{v_2}+{B_2}{\e}\cr
4{\l}{v_1}{v_3}+{\m}{\e} &
-4{\l}{v_3}{v_2}+{B_2}{\e} & {m_{{\tilde {\n}}_{\tau}}^2}
+2{\l}c+4{\l}v_{3}^2\cr}
\right)
\end{equation}

In a similar way, the neutral pseudoscalar mass-squared matrix becomes
$3 \times 3$ in the basis $\{Im(H_1), Im(H_2), Im(\tilde \nu_\tau)\}$ 
and takes the form:
\begin{equation}
M_{p}^2 = \left(\matrix
{m_{1}^2+2{\l}c & -{B_1}{\m} &
{\m}{\e}\cr
-{B_1}{\m} & m_{2}^2-2{\l}c &
-{B_2}{\e}\cr
{\m}{\e} &
-{B_2}{\e} & {m_{{\tilde {\n}}_{\tau}}^2}+2{\l}c\cr}
\right)
\end{equation}

In the charged scalar sector the charged scalar mass-squared matrix
becomes $4 \times 4$ in the basis $\{H_2, H_1, \tilde \tau_L, \tilde
\tau_R\}$ and is given by
\begin{equation}
{M_c}^2 = \left(\matrix
{r - {\frac{1}{4}}{g'}^2 c &
-{B_1}{\m} + {\frac{1}{2}}g^2 {v_1} {v_2} & -{B_2}{\e} + {\frac{1}{2}}
g^2 {v_2} {v_3} & -{\e} f {v_1}\cr
-{B_1}{\m} + {\frac{1}{2}}g^2 {v_1} {v_2} &
s + {\frac{1}{4}}{g'}^2 c + f^2 v^2_3 &
{\m}{\e} + {\frac{1}{2}}g^2 {v_1}{v_3} - {\frac{1} {2}}f^2 v_1v_3 & 
-{\e} f {v_2} - A f {v_3}\cr
-{B_2}{\e} + {\frac{1}{2}} g^2 {v_2} {v_3} &
{\m}{\e} + {\frac{1}{2}}g^2 {v_1}{v_3} - {\frac{1} {2}}f^2 v_1v_3 &
p + {\frac{1}{4}} g^2 t + {\frac{1}{4}}{g'}^2 c  &
{\m} f {v_2}+A f {v_1}\cr
-{\e} f {v_1} & -{\e} f {v_2} - A f {v_3} & {\m} f {v_2}+A f
{v_1} &
q -{\frac{1}{2}} {g'}^2 c + f^2 {v_3}^2\cr}
\right)
\end{equation}
with

$$r = m_{2}^2 + {\frac 1 4}{g^2}(v_{1}^2 + v_{2}^2 + v_{3}^2)$$
$$s = m_{1}^2 + {\frac 1 4}{g^2}(v_{1}^2 + v_{2}^2 - v_{3}^2)$$
$$p = m_{\tilde L}^2 + f^2 v_{1}^2$$
$$q = m_{\tilde R}^2 + f^2 v_{1}^2$$
$$t = (-v_{1}^2 + v_{2}^2 + v_{3}^2)$$
$$c = (v_{1}^2 -v_{2}^2 + v_{3}^2)$$
$$\l = (g^2 + {g'}^2)/8$$

In addition to constraints coming from 
electroweak symmetry breaking, the requirement of electric charge conservation
subjects the parameters in the potential to appropriate conditions in 
any given scenario. 

In the calculations here, we make the following choice of independent
parameters:

$\{\mu, {\rm {tan} \beta}, \epsilon, v_3, m^2_{\tilde \tau_L} =
m^2_{\tilde \tau_R} = m^2_{\tilde \nu_\tau} = m^2_0, A_t, A_b, A_{\tau},
B_1\}$. We also take the ramaining scalar mass parameters to be as follows:

$m^2_{\tilde {t_1}} = m^2_{\tilde {t_2}} = m^2_{\tilde {b_1}} =
m^2_{\tilde {b_2}} = m^2_{\tilde q}$

and 

$m^2_{\tilde e, \tilde \mu} = m^2_0$

where, ${\rm {tan} \beta}$ is the ratio of two Higgs vacuum expectation
values, $B_1$ is the usual B-term corresponding to the $\mu$-term, A's
are the trilinear soft breaking terms.

The bilinear R-parity violating parameters $\epsilon$ and $v_3$ can 
be constrained from different experimental considerations such as
the tau-mass measured within the existing errors, and, most importantly,
from the constraints on neutrino masses. The latter can be subjected to
current laboratory bounds, but the restriction becomes more stringent
on using the Superkamiokande results on atmospheric neutrinos. In this light, 
the values of $\epsilon$ and $v_3$ have to be constrained in such a way that
the sneutrino vev is less than a few hundred KeV's in the basis where
$\epsilon$ is rotated away. \footnote {Of course, bilinear terms $L_iH_2$ 
with $i=2,3$ have to be included in
order to explain $\nu_\mu - \nu_\tau$ oscillation. This requires a
straightforward extension of the formalism described above. For more
details, the reader is referred to Ref. \cite{17}.} 

The above kind of mixing in the chargino, neutralino and scalar sectors
results in physical states which are superpositions of 
neutral(charged)  leptons and neutralinos(charginos) on one hand,
and Higgses and sleptons(sneutrinos) on the other. This implies that
the Yukawa and gaugino coupling terms, written in terms of the 
physical states, will give rise to {\it all} the interactions of the  
$\lambda$ and $\lambda'$-types. In addition, the fact that gaugino
couplings play a role here makes it possible to free some of the terms
from constraints due to  gauge invariance of
the superpotential. One such interaction  term has particular importance
in our calculation, namely one involving a top quark, a stop and a
state which is dominantly a neutrino. This is clearly a result of
neutrino-neutralino mixing and the quark-squark neutralino interaction
in the Lagrangian. For more details the reader is referred to
\cite{18}.  

Also, a scenario like this allows one to have tau-neutralino-W and
neutrino-neutralino-Z  couplings. This results in additional decay modes
of a neutralino which are characteristic of bilinear R-parity violation
\cite{18}.

\underline {Case (2).}  $W_{\not R} = \lambda_{ijk} L_i L_j E_k^c 
+ \lambda_{ijk}' L_i Q_j D_k^c$  

In such a case the requirement of gauge invariance implies that  
$\lambda_{ijk}$s are 
antisymmetric in the first two indices due to $SU(2)$ invariance but
$\lambda_{ijk}'$s are not subjected to such constraints. Thus we have 
nine $\lambda$ and twenty seven $\lambda'$-type couplings other than 
the MSSM  parameter space. These trilinear couplings can also be
constrained in various ways like, for example, lepton universality 
violation \cite{19},
neutrinoless double beta decay \cite{20}, majorana mass of neutrino
\cite{21}, flavor
changing neutral current processes \cite{22} etc. Their implications 
in various high energy collider experiments such as LEP \cite{23}, HERA
\cite{24} or
the Fermilab Tevatron \cite{25} have also been widely explored. 
In addition, they can be responsible 
for neutrino masses generated at the one-loop level \cite{26}. 

Before we end this section, it may be remarked that although constraints
on the R-violating parameters are frequently  talked about, in practice
these constraints will always have to stipulate a given set of
values of the R-conserving (MSSM) parameters.  

\section{The one-loop calculations}

The one-loop calculation of the decay $\tilde \chi^0_1(p) \rightarrow
\nu(k_1) + \gamma(k_2)$ has been performed in the nonlinear R-gauge. A similar 
calculation for the radiative decay of heavier neutralinos into lighter ones 
can be found  in Ref. \cite{27}. In this gauge, 
choice of the gauge fixing term modifies certain types of vertices compared to,
say, the 't Hooft-Feynman gauge.  
For example, the $W^+G^-\gamma$ vertex (where $G^{\pm}$ is the charged 
Goldstone boson) is abesnt in the non-linear R-gauge, whereas the
$W^+W^-\gamma$ vertex gets correspondingly modified. 
Also a set of diagrams involving a $Z-\gamma$ transition (with an off-shell 
Z), which potentially contributes to the process under investigation, 
gives zero contribution when one sums over all the loops.  

The diagrams which ultimately contribute in the bilinear 
R-parity violating scenario are the triangle diagrams shown in Fig. 1.
A subset of these will be present when one considers only trilinear 
lepton number violating terms in the superpotential, the relevant
diagrams being  (a) and (b) involving
only the down-type quark/squark or lepton/sleptons in the loop. This is
actually a consequence of  $SU(2)$ invariance of the superpotential 
and can be seen
explicitly if one expands the $\lambda$- and $\lambda'$-type terms into
their component forms. The appearance of  additional diagrams in the
bilinear R-parity violating case can be attributed to mixing in
the scalar and lepton-chargino-neutralino sectors, once electroweak symmetry
breaking takes place (see section 2). In fact, as has already been mentioned, 
one of the main reasons for undertaking this study with bilinear R-parity 
violation  is the presence of the additional diagrams involving the top quark
and the W-boson, which are potential sources of enhancement of the
decay amplitude.
\input{mixed1.fig}

Since the photon is on-shell, $U(1)_{EM}$ gauge invariance demands
that the transition amplitude be proportional to $\sigma^{\mu\nu}
k_{2\mu} \epsilon^*_\nu.$ Thus one expects the following form for the
matrix element: 
\begin{equation} 
{\cal M} = i g_{{\tilde \chi^0_1} \nu \gamma} {\bar u}(k_1) 
(P_R - \eta_\nu \eta_1 P_L) \sigma^{\mu \nu}
k_{2\mu} \epsilon^*_\nu u(p)
\end{equation}
where $M_{\tilde \chi^0_1}$ is the mass of $\tilde \chi^0_1$, 
$\eta_1$ and $\eta_\nu$ are the signs of the mass eigenvalues of $\tilde
\chi^0_1$ and the neutrino respectively. Thus, depending on $\eta_1$ and
$\eta_\nu$, the effective $\tilde \chi^0_1 \nu \gamma$ interaction is
either proportional to $\sigma^{\mu \nu} k_{2\mu} \epsilon^*_\nu$ or
$\gamma_5 \sigma^{\mu \nu} k_{2\mu} \epsilon^*_\nu$. The radiative decay
width of $\tilde \chi^0_1$ is then given by
\begin{equation}
\Gamma(\tilde \chi^0_1 \rightarrow \nu \gamma) = \frac {g^2_{{\tilde
\chi^0_1} \nu \gamma} M^3_{\tilde \chi^0_1}} {16 \pi}
\end{equation}
where we have neglected the mass of the neutrino and $g^2_{{\tilde
\chi^0_1} \nu \gamma}$, containing details of the loop integrals, has 
the dimension of inverse mass-squared. 

In order to calculate $g_{{\tilde \chi^0_1} \nu \gamma}$ one has to
evaluate the triangle graphs.  For a given set of internal particles 
there are two diagrams which differ from each other by the direction of 
charge flow in the loop. In the figures we have shown only one set of 
diagrams. One must be careful about the diagrams involving the reverse 
flow of charge in the loop because in those cases one encounters vertices 
with clashing arrows.  

The integrals which appear during the loop calculations are regularized
using dimensional regularization. Since there is no tree-level coupling
of the type $\tilde \chi^0_1 \nu \gamma$ the divergences must cancel
among each gauge invariant subset of diagrams. For example, the diagrams
(g) and (h) and their partner graphs where the direction of charge flow
is reversed form a gauge invariant subset and it can be shown that the
infinities cancel within this set of diagrams. Similarly, one gets a
finite result out of the loop diagrams involving
fermion/sfermion loops, i.e., the diagrams $\{(a), (b)\}$
and their partner graphs. This is true, pairwise, for the loops
$\{(c),(d)\}$ and $\{(e),(f)\}$ involving charged Higgs boson and
charginos and the charged Goldstone bosons and the
charginos.

The decay amplitude shown in Eqn.(13) can be decomposed in the
following way (with only the appropriate contributions retained,  as 
stated earlier, with trilinear R-violating terms);

\begin{equation}
  {\cal M} =    {\cal M}_{1} + {\cal M}_{2}
\end{equation}

The matrix element due to the four diagrams (a), (b) and their partner
graphs is given by

\begin{eqnarray}
{{\cal M}_{1}}  &=&  {\frac {e g^2 \eta_\nu}{16 \pi^2}}{\bar u}(k_1)
(P_R - \eta_\nu \eta_1 P_L){{\not k}_2} {\not \epsilon}^*u(p) \nonumber
   \\[1.5ex]
& {\times} & \sum_{f} e_f C_f \{(A_LB_R - A_RB_L)
[\eta_1 {M_{\tilde \chi^0_1}} (I^1 - I^3)]  +  m_f (A_LB_L - A_RB_R)I^2\}
\end{eqnarray}

\noindent
where
$Q_f$ is the charge of the fermion $f$ in units of e $(e > 0)$, and
$C_f$ is the color factor for the particles in the loop. One can use
this result to include contributions from all possible fermion/scalar
loops. Thus contributions from charged Higgs and Goldstone
bosons are also present in the sum over $f$. We have also explicitly used 
the sum over the three possible chargino states
($\tilde \chi^+_k$) in the bilinear R-parity violating scenario.
One must be careful with the
masses and couplings as well as about the electric charges of the
fermions in the loop. The integrals $I^i$ are expressible 
in terms of the
general expressions for one-loop three-point functions \cite{28}, in forms 
that are presented in the appendix, where we also give the expressions 
for the combinations $(A_LB_L - A_RB_R)$ and $(A_LB_R - A_RB_L)$ for 
all the possible cases. 

In the case of stop and sbottom we have included
the effect of left-right mixing. Also, the calculation includes possible
absorptive parts of the loop integrals, which can be present in particular
when the decaying neutralino is heavier than the $W$.  

In a similar way, the contributions to the matrix element from 
diagrams (g) and (h) and their partner graphs are given by

\begin{eqnarray}
{{\cal M}_{2}}  &=&  {\frac {-e g^2 \eta_\nu}{4 \pi^2}}{\bar u}(k_1) 
(P_R - \eta_\nu \eta_1 P_L){{\not k}_2} {\not \epsilon}^*u(p) \nonumber
   \\[1.5ex]
& {\times} & \sum_{k} \{(A_LB_L - A_RB_R)
[\eta_1 {M_{\tilde \chi^0_1}} (I^1 - J^2 -I^3)]  + 2 M_k (A_LB_R -
A_RB_L)J^2\} 
\end{eqnarray}

In both the above expressions, the quantities $A_L$, $A_R$, $B_L$ and $B_R$ 
are all assumed to be real.

The loops involving the top quark (Fig.1(a) and Fig.1(b)) , which are 
present only in the case of bilinear R-parity violation, always seem 
to have a rather important effect \footnote {Our detailed scan of the 
parameter space, however, also reveals regions where lighter fermions 
contribute comparably because of favoured mixing angles, much in the same 
way as the charm quark makes dominant contributions to the real part of the 
box diagrams for $K^0-{\bar K}^0$ mixing.}.
Matching contributions also come over a large area of the parameter space
from loops involving the W-boson (and the tau and/or the lighter chargino,
depending on the relevant mixing angles) in the propagator 
(diagrams (g) and (h)). 
The very presence of these diagrams 
causes the rates for radiative decays to be larger in cases with bilinear
R-violating effects than in those with only trilinears, except in those
where there is destructive interference between the two types of graphs.

\section{Numerical Results}
The calculation of the branching ratio for
$\tilde \chi^0_1 \rightarrow \nu \gamma$ involves the computation of 
decay widths of the lightest neutralino in all the relevant
tree-level two-and three-body decay modes. The different types of mixing 
that have been discussed in section 2 for  bilinear R-parity violation
open up the following two-body decay modes, as and when
kinematically allowed:

$\tilde \chi^0_1 \rightarrow \tau^{\mp} W^{\pm},
\tilde \chi^0_1 \rightarrow \nu_\tau Z,
\tilde \chi^0_1 \rightarrow \nu_\tau h$
where h is the lightest Higgs boson.

In addition,  there are three body tree-level decay channels with  both
bilinears and trilinears. While the channels 
$\tilde \chi^0_1 \rightarrow e^+_i {\bar u}_j d_k, 
\tilde \chi^0_1 \rightarrow  \nu_i {\bar d}_j d_k, 
\tilde \chi^0_1 \rightarrow e^+_i \nu_j e^-_k$ 
are available in both the cases \cite{29},  additional decays such as 
$\tilde \chi^0_1 \rightarrow u_i \bar{u_i} \nu$ and
$\tilde \chi^0_1 \rightarrow \nu \nu \nu$ 
can take place in  the former, and can be important when two-body 
decays are not kinematically allowed.

In all the numerical results presented here, we have used 
$A_t = A_b = 10 GeV$. Our finding is that the
branching ratio of the radiative  decay is rather insensitive to the
values of these parameters.

Let us first concentrate on bilinear R-parity violating scenario. We
will consider two different cases, namely: (i) when the supersymmetry breaking
gaugino mass parameters are unified at the 
Grand Unification scale, and (ii) when 
$SU(2)$ and $U(1)$ gaugino mass parameters denoted by $M_2$ and $M_1$
respectively are treated as unrelated.  

Due to reasons explained below, significant contributions to the
radiative decay can come only when bilinear R-parity violating effects are 
present. There again, the relevant parameters $\epsilon$ and the sneutrino
vev(s) can be constrained under two different considerations. If the
neutrino-neutralino mixing process in such a scenario is the main 
mechanism for the generation of (Majorana) neutrino masses for explaining
the Superkamiokande (SK) results on atmospheric muon neutrinos, then
the parameters have to be constrained in such a way that the quantity 
$v' = \sqrt{v^2_{\mu} + v^2_{\tau}}$ is less than about 100 KeV {\it in a
basis where the $\epsilon$-parameters are rotated away from the
superpotential}.  The numerical results modulo such constraints
are denoted by ``with SK'' in the corresponding figure captions, and the
results correspond to both the second-and third-generation neutrinos in
the final state of the radiative decays. On the other 
hand, if one ignores the SK constraints, then, with just the 
laboratory bounds on the tau neutrino mass, the values of $\epsilon$ and
$v_3$ (i.e. the R-violating parameters in the third generation) can be
as large as on the order of GeV's. We have also presented such results, 
considering only  the third (tau) neutrino to be there as the decay
product.

Figures 2-9 (10) show the results of a scan of the parameter space carried
out in the different scenarios mentioned above, calculated with  
bilinear (trilinear) R-parity violation. The first thing that we note 
is that in all the results presented, a specific set of masses for the
squarks and sleptons are assumed. The branching ratios of the radiative decay,
however, are rather insensitive to their variation. This is because if,
for example, we reduce the squark masses, the effect of a lighter stop
will increase the decay width for the radiative decay, but the corresponding
tree level decays will also undergo a boost via diagrams mediated by
squarks. 

Similarly, a comparison of figures 2-4 with 5-7 and 8 with 9 reveals that
imposing the SK constraints does not cause any appreciable reduction
to the probability of obtaining branching ratios on the higher side
for similar combinations of MSSM parameters. All that it does is to
scale the overall coefficients instrumental in both the loop-and
tree-level decays. This results in branching ratios of similar magnitudes,
but is manifested in larger decay lengths for the neutralino when
R-parity violating parameters are restricted to yield
$\Delta m^2_{23} \sim a~~ few ~~times~~ 10^{-3}$. 
 
As has already been mentioned, the major contributions to the radiative
decay come from {\it(a)} the top-stop loop with bilinear R-pariy violation,
as well as from loops involving the W-boson and the tau or the lighter 
chargino. However, this leads in some regions of the parameter space
to the interesting possibility of their cancelling each other. When such a
cancellation is very severe for some specific neutralino mass, the
branching ratio is seen to undergo a sharp dip at that point, as seen
in Fig.3 and Fig.8. In such cases, the effective contributions
at those points hardly contain any input that is special of the
bilinear terms. On the other hand, in Fig.2 (and partially in Fig.3) 
one encounters a situation where the contribution is mainly from the 
W-loops but the latter undergoes a destructive interference among the 
various component terms, causing the overall branching ratio to fall 
at a particular region.

The numerical results show a sensitivity to the MSSM parameters
$\mu$ and, to a somewhat lesser extent, on $\tan \beta$.
It is also clear from the graphs that the radiative decay tends to remain
suppressed for small values of $|{\mu}|$ which is treated here as a 
free parameter. It gradually rises with
$|{\mu}|$, and then almost saturates, showing a slight fall for 
$|{\mu}|$ approaching a TeV. The loops have the largest contributions 
for $|{\mu}| \sim 500 GeV$. By and large, the corresponding regions of the 
parameter space  have the lightest neutralino  almost entirely dominated
by the Bino state.

When the condition of gagino mass unification is relaxed, 
the mass parameters $M_1$ and $M_2$, corresponding to the $U(1)$ and $SU(2)$
gauginos can be unrelated \cite{30}. Most of the features of the unified
scenario, including the possibilities of having branching ratios close
to 10 per cent, are seen here also. However, it entails the 
additional possibility of having a destructive interference between the
top-and W-induced diagrams in cases where the branching ratio is otherwise
on the higher side, causing the latter to fall sharply by about 5 orders
of magnitudes for a particular value of the neutralino mass, with all
other parameters at same values. This is seen when  $M_2$/$M_1$ is smaller
than  what it would have been with the constraint of unification, and
the dip is found to occur when the two masses are quite close to each other.

Figure 10 is a sample showing the order of magnitude of the radiative decay
with only trilinear R-violating couplings present in the theory. The
values of the $\lambda$- and $\lambda'$-type coupling constants have
been used consistently with the current limits \cite{31}. We have
already seen that the two potentially most important classes of diagrams,
namely, those mediated by the top and the W,
are absent in such a case. Therefore, it is hardly surprising that the
branching ratios cannot rise above $10^{-8}$ - $10^{-10}$ over a large
part of the parameter space. 
The range in which the branching ratio tends to lie in such a case
matches, as one would expect,  with the one to which it falls when a 
cancellation of the large contributions takes place in the scenario 
with bilinears. In any case, the radiative branching ratio turns out to
be too small to be of any observable significance where R-parity is
violated only through trilinear terms.

With the maximum value of the branching ratio for the radiative decay
being between 5 and 10 per cent, two-photon signals from such decays
will fake signals like those of GMSB in only a rather small region of the
parameter space. Such a thing might happen  when the lighest neutralino
in GMSB is close to the kinematic limit of production. However, the
fact that only the bilinear R-violating terms can boost the
branching ratio to the level of close to 10 per cent has rather interesting
implications in terms of observing distinctive signals of the latter.
For example, pair-produced neutralinos at LEP energies (assuming an 
intergated luminosity of 500 $pb^{-1}$) can give rise to about 40
events where there is a radiative decay on one side, leading to 
signals of the type $3f + \gamma$. There can be many more of such events
in a high-energy electron-positron collider with an integrated
luminosity of about 50 $fb^{-1}$. With a proper event selection strategy,
radiative neutralino decay can thus be an interesting signature of
R-parity violation with bilinear terms.

\section{Summary and conclusions} 
 
We have performed a detailed calculation of the branching ratios for
the radiative decay $\tilde \chi_1^0 \rightarrow \nu\gamma$ for the lightest
neutralino in R-parity violating SUSY models. It is seen that
the branching ratio can have a maximum value of about 5-10\% when R-parity
violation has its origin in the so-called `bilinear' terms in the
superpotential, and is insignificantly small for that induced by
`trilinear' terms. This, we point out, can be an interesting way of
obtaining characteristic  signals for the former type of scenario. Our 
conclusion is  that the chances of such radiative decays faking the 
two-photon signals for gauge-mediated SUSY breaking are small, except
where the lightest neutralino in the latter is close to the
kinematic limit of production. It is
also seen that the accessible region corresponding to the MSSM
parameter space does not change appreciably upon subjecting
the theory to constraints from neutrino mass patterns  required by the
Superkamiokande data on atmospheric neutrinos.  

{\bf Acknowledgment:} We wish to acknowledge Aseshkrishna Datta and
Debajyoti Choudhury for useful discussions and help in preparing the
figures.

\newpage
\appendix
\renewcommand{\theequation}{A.{\arabic{equation}}}
\setcounter{equation}{0}
\bigskip

\bigskip
\noindent {\large {\bf Appendix}}\\
\noindent Here we outline the actual forms of the various quantities
used in the loop calculations in section 3. 

Expressions for the mixing parameters $A_L$, $B_L$, $A_R$, 
$B_R$ for $W^{\pm}$-loops in the bilinear R-parity violating scenario:
\begin{equation}
A_L = {\frac {1} {\sqrt{2}}}Z_{i1}V_{k2} - Z_{i3}V_{k1}
\end{equation}
\begin{equation}
B_L = {\frac {1} {\sqrt{2}}}Z_{j1}V_{k2} - Z_{j3}V_{k1}
\end{equation}
\begin{equation}
A_R = -{\frac {1} {\sqrt{2}}}Z_{i2}U_{k2} - Z_{i3}U_{k1} + {\frac {1}
{\sqrt{2}}}Z_{i5}U_{k3}
\end{equation}
\begin{equation}
B_R = -{\frac {1} {\sqrt{2}}}Z_{j2}U_{k2} - Z_{j3}U_{k1} + {\frac {1}
{\sqrt{2}}}Z_{j5}U_{k3}
\end{equation}
where we have j = 2 and i = 1.
Similarly for charged scalar loops we have:

\begin{eqnarray}
A_L &=& m'(Z_{i5}V_{k3}C_{2m} - Z_{i2}V_{k3}C_{3m}) + 2
Q_{\tau}\tan\theta_{W}Z_{i4}V_{k3}C_{4m} +\nonumber
  \\[1.5ex]
& &\sqrt{2}\{Z_{i1}V_{k1}
+ {\frac {1} {\sqrt{2}}}(Z_{i4}V_{k2}\tan\theta_{W} +
Z_{i3}V_{k3})\}C_{1m}
\end{eqnarray}

\begin{eqnarray}
A_R &=& -m'Z_{i2}U_{k3}C_{4m} + (Z_{i3} - Z_{i4}\tan\theta_{W}(1 + 2
Q_{\tau}))U_{k3}C_{3m} +\nonumber 
  \\[1.5ex]
& &\sqrt{2}\{Z_{i2}U_{k1}
- {\frac {1} {\sqrt{2}}}(Z_{i4}U_{k2}\tan\theta_{W} +
Z_{i3}U_{k3})\}C_{2m}
\end{eqnarray}

\begin{eqnarray}
B_L &=& -m'Z_{j2}U_{k3}C_{4m} + (Z_{j3} - Z_{j4}\tan\theta_{W}(1 + 2
Q_{\tau}))U_{k3}C_{3m} +\nonumber 
  \\[1.5ex]
& &\sqrt{2}\{Z_{j2}U_{k1}
- {\frac {1} {\sqrt{2}}}(Z_{j4}U_{k2}\tan\theta_{W} +
Z_{j3}U_{k3})\}C_{2m}
\end{eqnarray}

\begin{eqnarray}
B_R &=& m'(Z_{j5}V_{k3}C_{2m} - Z_{j2}V_{k3}C_{3m}) + 2
Q_{\tau}\tan\theta_{W}Z_{j4}V_{k3}C_{4m} +\nonumber
  \\[1.5ex]
& &\sqrt{2}\{Z_{j1}V_{k1}
+ {\frac {1} {\sqrt{2}}}(Z_{j4}V_{k2}\tan\theta_{W} +
Z_{j3}V_{k3})\}C_{1m}
\end{eqnarray}
where
\begin{equation}
m' = \frac {m_{\tau}} {(m^2_W - \frac{g^2 v^2_3}{2})^{\frac {1}
{2}}{\cos\beta}}
\end{equation}

\noindent Z's are the neutralino mixing elments, U's and V's are the
chargino mixing elements and C's are the charged scalar mixing elements. 

Special cases for stop and sbottom: we have considered the effect of
mixing between ${\tilde t}_L$ and ${\tilde t}_R$ as well as between
${\tilde b}_L$ and ${\tilde b}_R$ \\
(i) For $t$-${\tilde t}_1$ triangle:
\begin{equation}
A_L = (Z^-_1 + 2Q_tZ_{14}\tan\theta_W)(-\sin\theta_t)  
+ {\frac {1} {m_W \sin\beta}} m_t Z_{11}\cos\theta_t
\end{equation}

\begin{equation}
A_R = {\frac {m_t} {m_W \sin\beta}} Z_{11}(-\sin\theta_t) - 2 Q_t \tan\theta_W
Z_{14}\cos\theta_t
\end{equation}

\begin{equation}
B_L = {\frac {m_t} {m_W \sin\beta}} Z_{21}(-\sin\theta_t) - 2 Q_t \tan\theta_W
Z_{24}\cos\theta_t
\end{equation}

\begin{equation}
B_R = (Z^-_2 + 2Q_tZ_{24}\tan\theta_W)(-\sin\theta_t)  
+ {\frac {1} {m_W \sin\beta}} m_t Z_{21}\cos\theta_t
\end{equation}
In our convention, ${\tilde t}_1$ is the lightest physical stop and
couplings or ${\tilde t}_2$ can be obtained by replacing $-\sin\theta_t
\rightarrow \cos\theta_t$ and $\cos\theta_t \rightarrow
\sin\theta_t$

(ii) For $b$-${\tilde b}_1$ triangle:
\begin{equation}
A_L = (-Z^-_1 + 2Q_bZ_{14}\tan\theta_W)(-\sin\theta_b)  
+ {\frac {1} {m_W \sin\beta}} m_b Z_{12}\cos\theta_b
\end{equation}

\begin{equation}
A_R = {\frac {m_b} {m_W \sin\beta}} Z_{12}(-\sin\theta_b) - 2 Q_b \tan\theta_W
Z_{14}\cos\theta_b
\end{equation}

\begin{equation}
B_L = {\frac {m_b} {m_W \sin\beta}} Z_{22}(-\sin\theta_b) - 2 Q_b \tan\theta_W
Z_{24}\cos\theta_b
\end{equation}

\begin{equation}
B_R = (-Z^-_2 + 2Q_bZ_{24}\tan\theta_W)(-\sin\theta_b)  
+ {\frac {1} {m_W \sin\beta}} m_b Z_{22}\cos\theta_b
\end{equation}
with 

\begin{equation}
Z^-_1 = Z_{13} - Z_{14}\tan\theta_W
\end{equation}

\begin{equation}
Z^-_2 = Z_{23} - Z_{24}\tan\theta_W
\end{equation}
In our convention, ${\tilde b}_1$ is the lightest physical sbottom and
couplings or ${\tilde b}_2$ can be obtained by replacing $-\sin\theta_b
\rightarrow \cos\theta_b$ and $\cos\theta_b \rightarrow
\sin\theta_b$

(iii) Degenerate sfermion loops:
For up-type squarks we can use the same expression as that for the stop 
loop without the mixing angle corresponding to ${\tilde t}_L$ - ${\tilde
t}_R$ mixing. Similarly for down-type squark or charged sleptons, one can
use the expressions corresponding to the sbottom loop modulo the mixing
factor. Also the charges and the color factors should be properly
included.

\noindent In the case of $\lambda$-type couplings the expressions become
(for the down-type squarks):
\begin{equation}
A_L = 0(\lambda')
\end{equation}

\begin{equation}
A_R = \lambda(0)
\end{equation}

\begin{equation}
B_L = {\frac {m_d} {m_W \cos\beta}}Z_{12} - 2Q_dZ_{14}\tan\theta_W
\end{equation}

\begin{equation}
B_R = {\frac {m_d} {m_W \cos\beta}}Z_{12} + 2Q_dZ_{14}\tan\theta_W -
Z^-_1
\end{equation}
where appropriate generational indices should be considered for the
$\lambda$- and $\lambda'$-type couplings.

\noindent In the case of charged sleptons one should use the appropriate
masses and charge. One must remember that in the case of trilinear
R-parity violating scenario we have diagonalised a $4 \times 4$
neutralino mass matrix.

The integrals corresponding to the three-point functions are
\begin{equation}
C_0 = {\frac {1} {i\pi^2}} \int {d^4q \frac {1} {(q^2 + m^2_f) (q -
p)^2 + m^2_s) ((q - k_2)^2 + m^2_f)}}   
\end{equation}
\begin{equation}
C_\mu = {\frac {1} {i\pi^2}} \int {d^4q \frac {q_\mu} {(q^2 + m^2_f) (q -
p)^2 + m^2_s) ((q - k_2 )^2 + m^2_f)}}   
\end{equation}
\begin{equation}
C_{\mu\nu} = {\frac {1} {i\pi^2}} \int {d^4q \frac {{q_\mu}{q_\nu}} {(q^2 
+ m^2_f) (q - p)^2 + m^2_s) ((q - k_2 )^2 + m^2_f)}}   
\end{equation}
They are expressed in terms of various form-factors as follows
\begin{equation}
C_\mu = -p_{\mu} C_{11} + k_{1\mu} C_{12}
\end{equation}
\begin{equation}
C_{\mu\nu} = p_{\mu} p_{\nu} C_{21} + k_{1\mu} k_{1\nu} C_{22} - (p_\mu
k_{1\nu} + p_\nu k_{1\mu})C_{23} + \delta_{\mu\nu} C_{24}
\end{equation}
and 
\begin{equation}
k_{1\mu} = p_{\mu} - k_{2\mu}
\end{equation}

\begin{equation}
I^1 = C_{12} - C_{11}
\end{equation}

\begin{equation}
I^2 = C_0
\end{equation}

\begin{equation}
I^3 = I^4 + I'^4
\end{equation}

\begin{equation}
I^4 = C_{23} - C_{22}
\end{equation}

and $I'^4$ can be obtained from $I^4$ by $m_s \leftrightarrow m_f$
interchange. For more details on the three-point form-factors, see, ref.
\cite{28}.

$J^2$ is analogous to $I^2$ but the denominator in Eqn.[A.24] will be 
changed to 
$$(q^2 + m^2_f)\{(q - k_1)^2 + m^2_s\}\{(q -p)^2 + m^2_s\}$$
The integrals which appear in the calculations of ${\cal M}_{2}$ are
identical in form to those mentioned above but one should replace $m_s$
with $m_W$ and $m_f$ with $M_k$. Here $m_s$ and $m_f$ are the masses of
the scalar and the fermion appeared in the loop respectively and $M_k$
is the mass of the chargino.
 
\begin{figure}[hbt]
\centerline{\epsfig{file=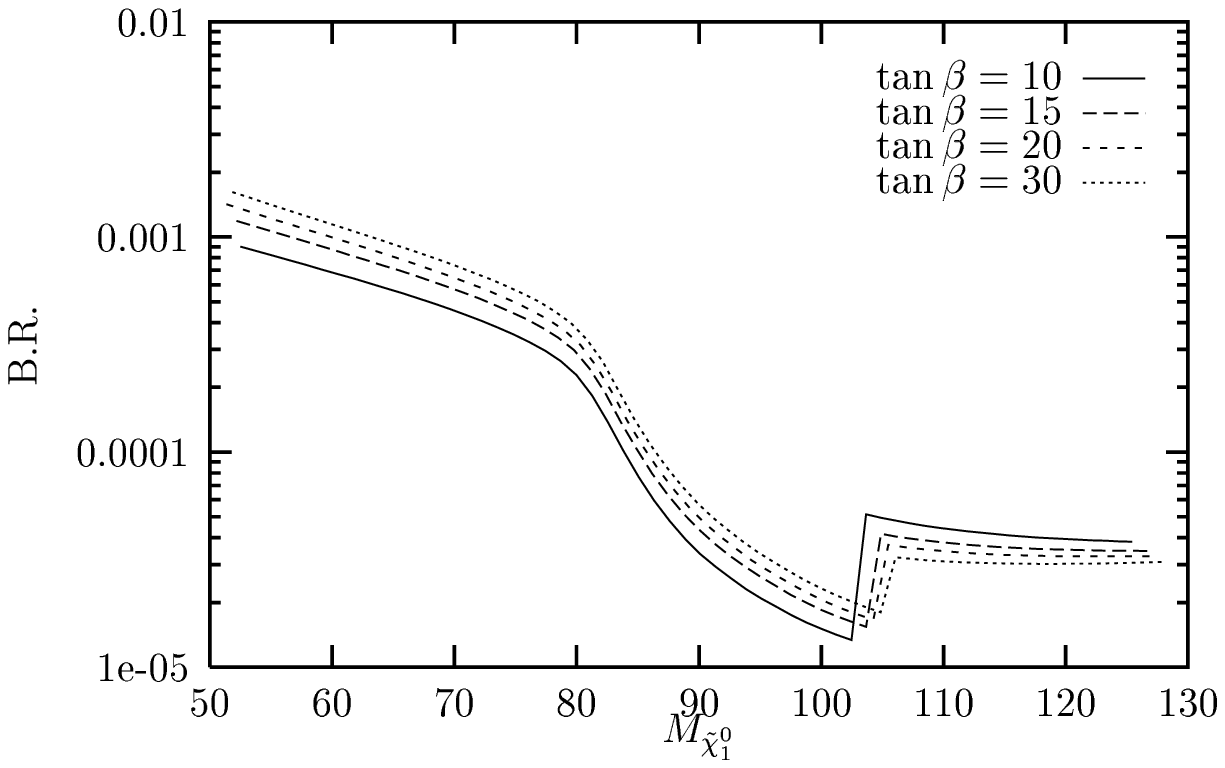,width=10cm}}
\caption{Branching ratio for
the decay $\tilde \chi_1^0 \rightarrow \nu\gamma$ (with SK).
The remaining supersymmetric parameters are chosen as:
$\mu=200$ GeV, $B_1 = -200$ GeV, $m_{\tilde
l} = m_{\tilde \nu} = 200$ GeV, $m_{\tilde q} = 600$ GeV.}
\end{figure}

\begin{figure}[hbt]
\centerline{\epsfig{file=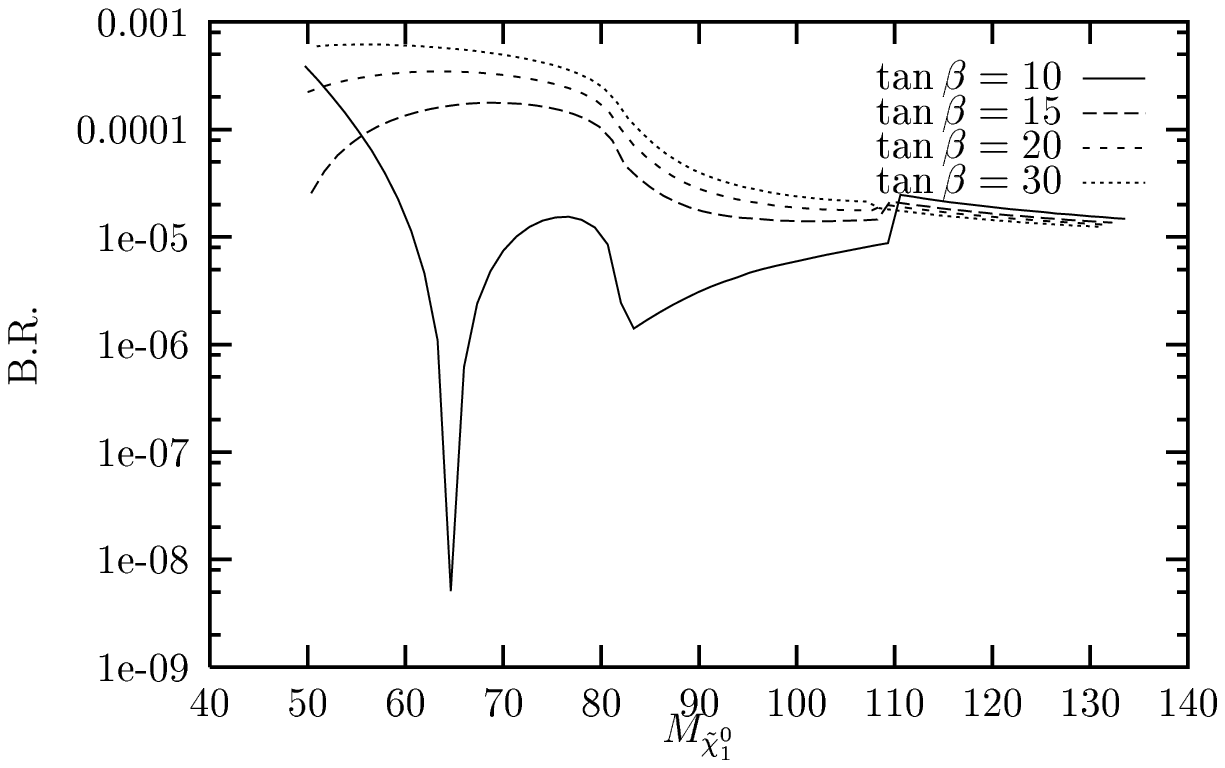,width=10cm}}
\caption{Branching ratio for
the decay $\tilde \chi_1^0 \rightarrow \nu\gamma$ (with SK).
The remaining supersymmetric parameters are chosen as:
$\mu=-200$ GeV, $B_1 = 200$ GeV, $m_{\tilde
l} = m_{\tilde \nu} = 200$ GeV, $m_{\tilde q} = 600$ GeV.}
\end{figure}

\begin{figure}[hbt]
\centerline{\epsfig{file=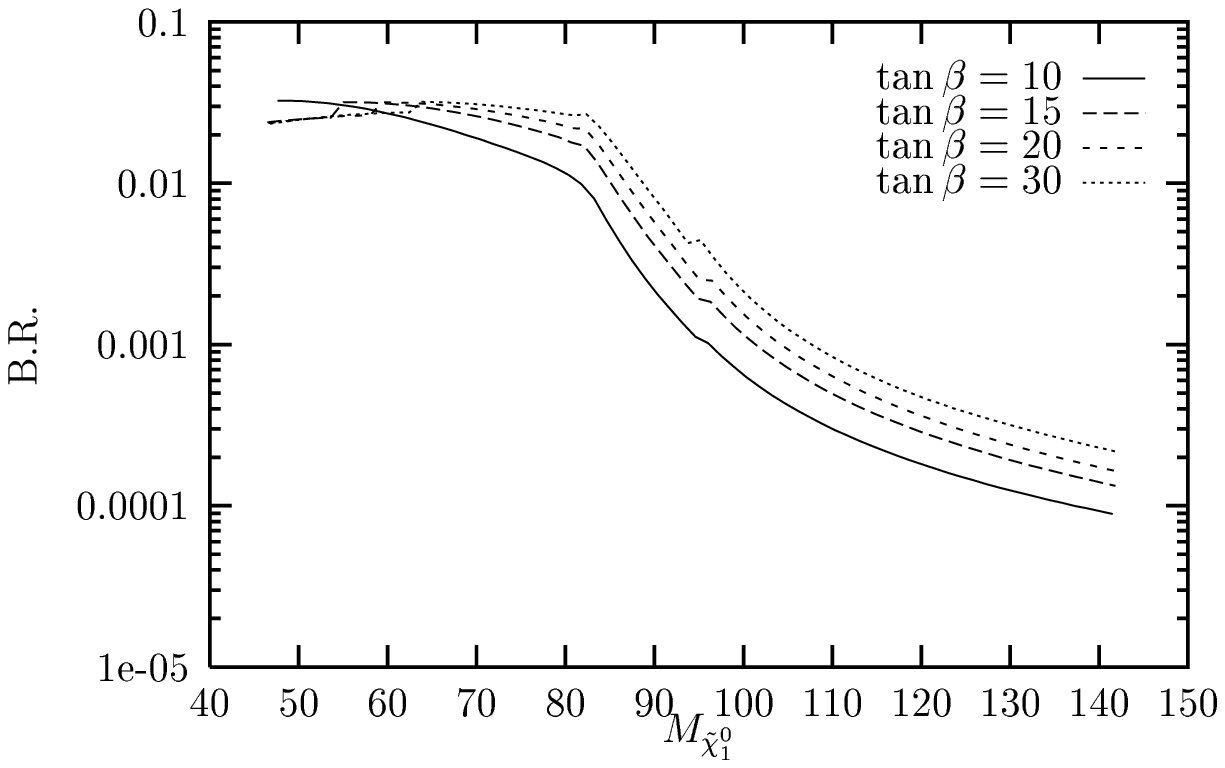,width=10cm}}
\caption{Branching ratio for
the decay $\tilde \chi_1^0 \rightarrow \nu\gamma$ (with SK).
The remaining supersymmetric parameters are chosen as:
$\mu=500$ GeV, $B_1 = -200$ GeV, $m_{\tilde
l} = m_{\tilde \nu} = 200$ GeV, $m_{\tilde q} = 600$ GeV.}
\end{figure}

\begin{figure}[hbt]
\centerline{\epsfig{file=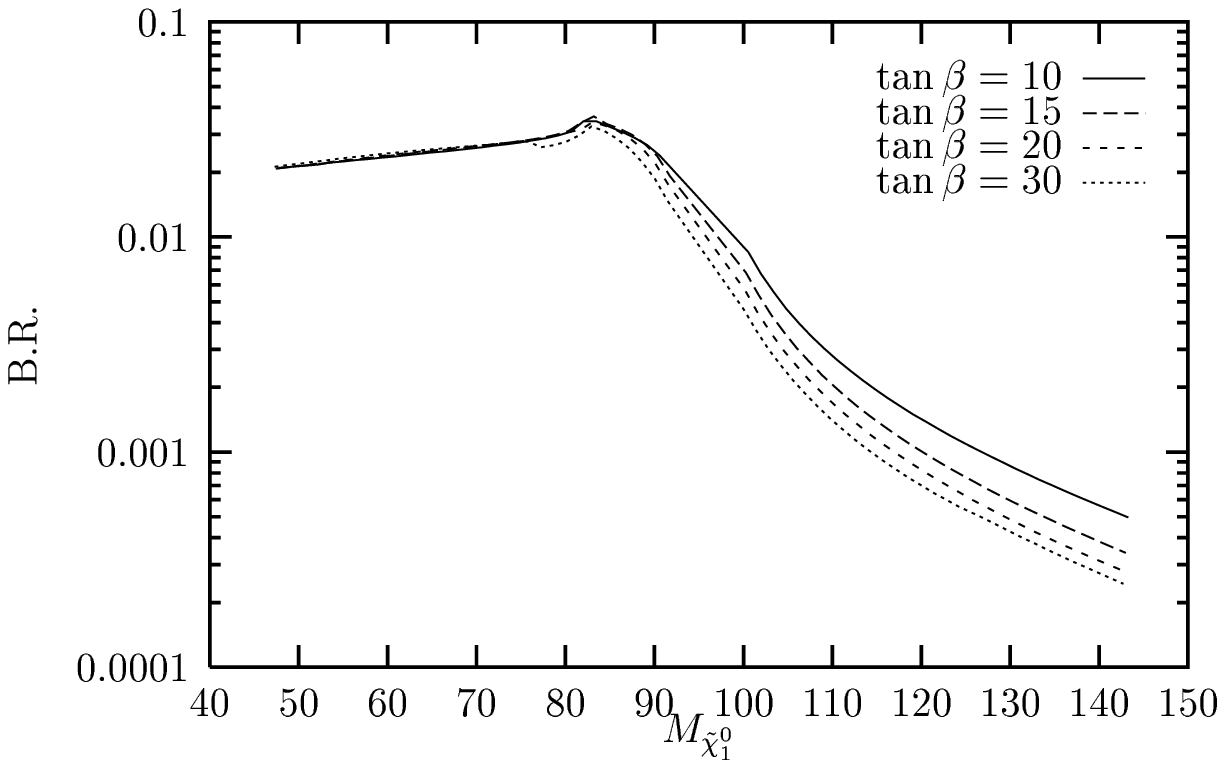,width=10cm}}
\caption{Branching ratio for
the decay $\tilde \chi_1^0 \rightarrow \nu\gamma.$
The remaining supersymmetric parameters are chosen as:
$\mu=-500$ GeV, $\epsilon = 10$ GeV, $v_3 = 1$ GeV, $B_1 = 200$ GeV, 
$m_{\tilde l} = m_{\tilde \nu} = 200$ GeV, $m_{\tilde q} = 600$ GeV.}
\end{figure}

\begin{figure}[hbt]
\centerline{\epsfig{file=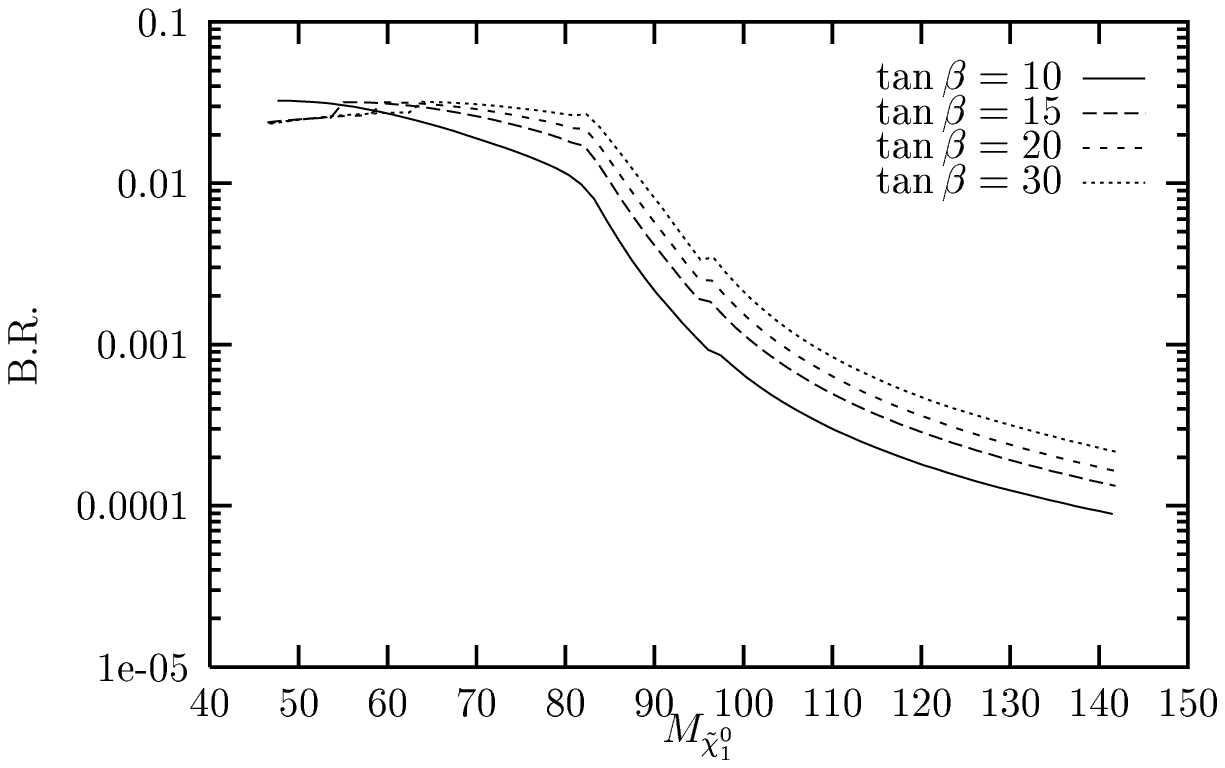,width=10cm}}
\caption{Branching ratio for
the decay $\tilde \chi_1^0 \rightarrow \nu\gamma.$
The remaining supersymmetric parameters are chosen as:
$\mu=500$ GeV, $\epsilon = 10$ GeV, $v_3 = 1$ GeV, $B_1 = -200$ GeV, 
$m_{\tilde l} = m_{\tilde \nu} = 200$ GeV, $m_{\tilde q} = 600$ GeV.}
\end{figure}

\begin{figure}[hbt]
\centerline{\epsfig{file=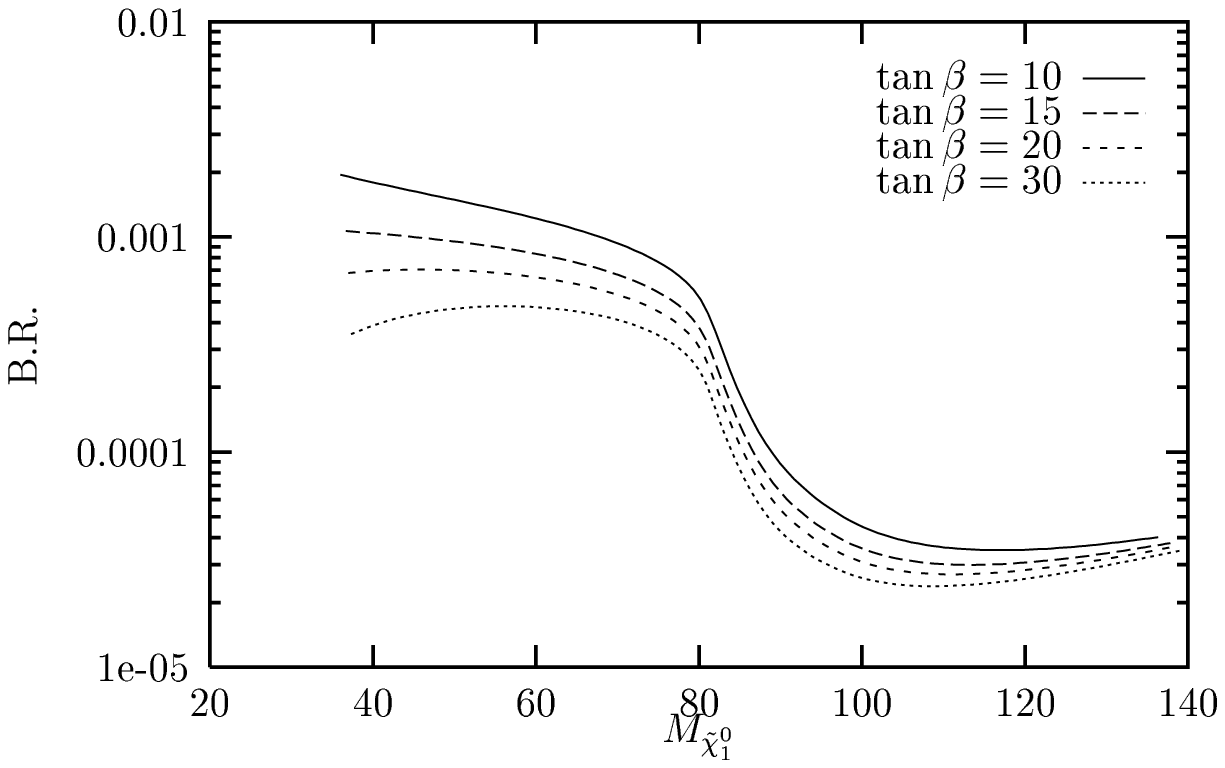,width=10cm}}
\caption{Branching ratio for
the decay $\tilde \chi_1^0 \rightarrow \nu\gamma$ (with SK), where 
$M_1$ and $M_2$
are free parameters. The value of $M_2$ is taken to be 300 GeV.
The remaining supersymmetric parameters are chosen as:
$\mu=200$ GeV, $B_1 = -200$ GeV, $m_{\tilde
l} = m_{\tilde \nu} = 200$ GeV, $m_{\tilde q} = 600$ GeV.}
\end{figure}

\begin{figure}[hbt]
\centerline{\epsfig{file=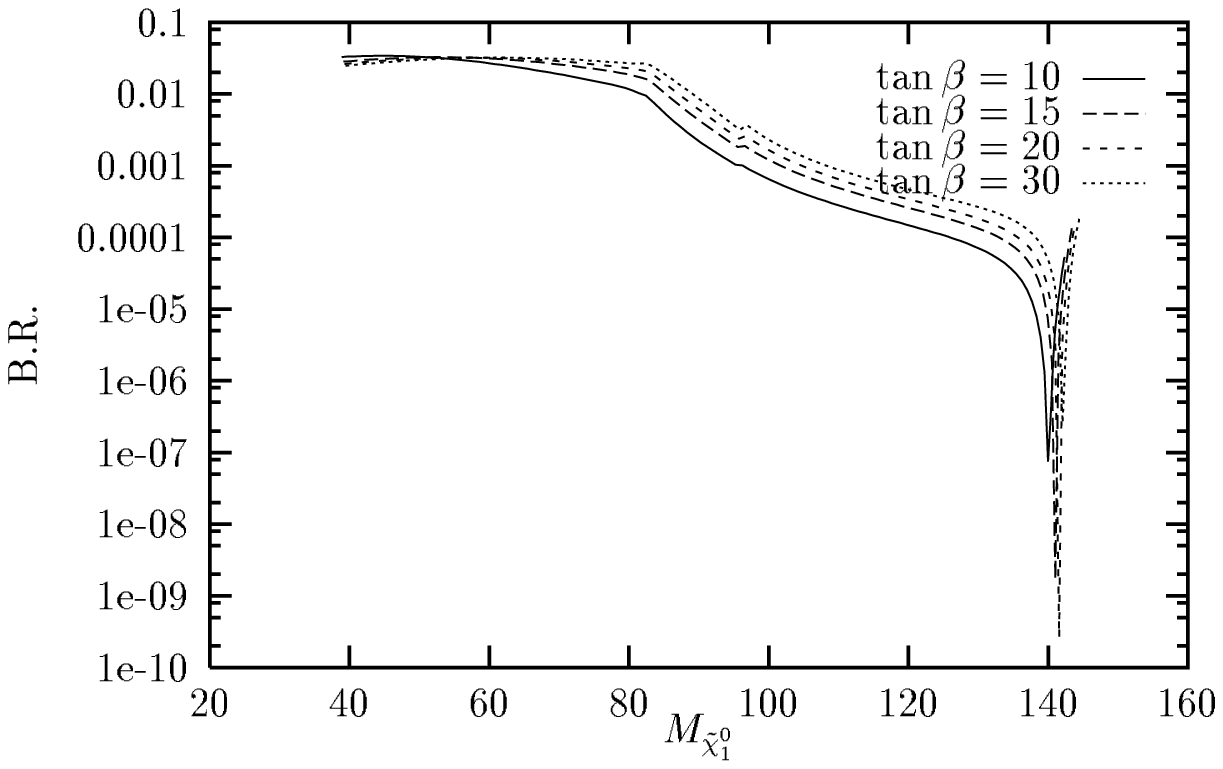,width=10cm}}
\caption{Branching ratio for
the decay $\tilde \chi_1^0 \rightarrow \nu\gamma$ (with SK), where 
$M_1$ and $M_2$
are free parameters. The value of $M_2$ is taken to be 150 GeV.
The remaining supersymmetric parameters are chosen as:
$\mu=500$ GeV, $B_1 = -200$ GeV, $m_{\tilde
l} = m_{\tilde \nu} = 200$ GeV, $m_{\tilde q} = 600$ GeV.}
\end{figure}

\begin{figure}[hbt]
\centerline{\epsfig{file=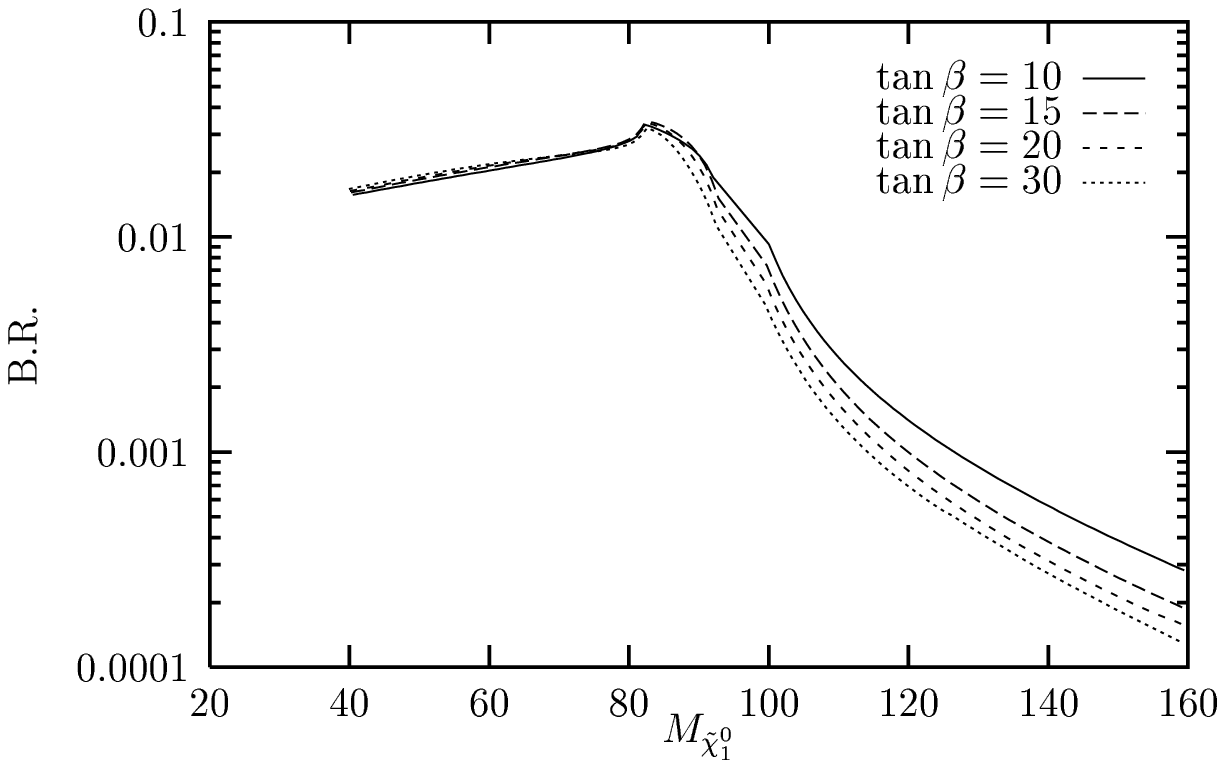,width=10cm}}
\caption{Branching ratio for
the decay $\tilde \chi_1^0 \rightarrow \nu\gamma$, where $M_1$ and $M_2$
are free parameters. The value of $M_2$ is taken to be 300 GeV.
The remaining supersymmetric parameters are chosen as:
$\mu=-500$ GeV, $\epsilon = 10$ GeV, $v_3 = 1$ GeV, $B_1 = 200$ GeV, 
$m_{\tilde l} = m_{\tilde \nu} = 200$ GeV, $m_{\tilde q} = 600$ GeV.}
\end{figure}

\begin{figure}[hbt]
\centerline{\epsfig{file=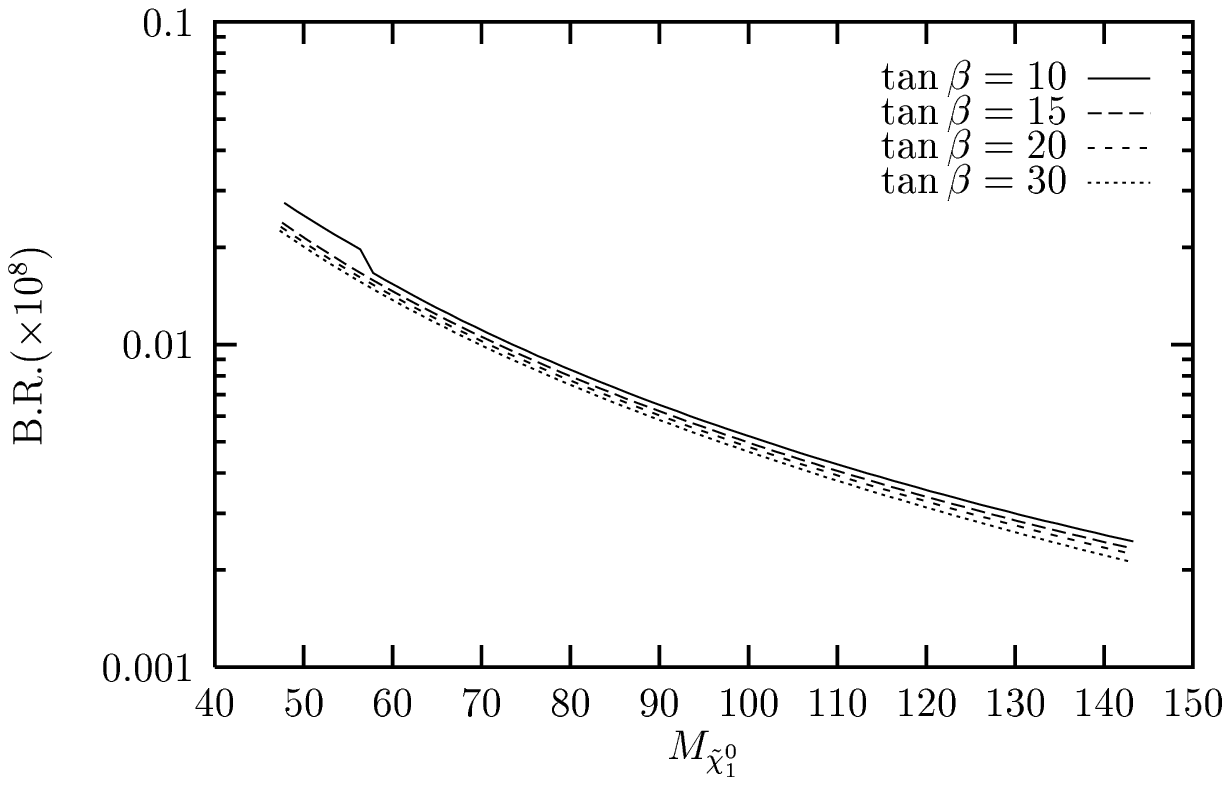,width=10cm}}
\caption{Branching ratio for
the decay $\tilde \chi_1^0 \rightarrow \nu\gamma$ in the scenario where
only $\lambda$ and $\lambda'$ terms are present.
The remaining supersymmetric parameters are chosen as:
$\mu=-500$ GeV, $m_{\tilde l} = 200$ GeV, $m_{\tilde d} = 600$ GeV.}
\end{figure}

\end{document}